\documentclass[aps,prl,twocolumn,groupedaddress,amsmath,amssymb]{revtex4-1}
\usepackage{ae}
\usepackage[pdftex]{graphicx}
\usepackage{amsmath}
\usepackage{amssymb}
\usepackage{epsfig, float}
\usepackage{hyperref}

\begin{document}


\title{Internal character dictates phase transition dynamics between \\ isolation and cohesive grouping}


\author{Pedro D. Manrique$^1$, Pak Ming Hui$^2$, Neil F. Johnson$^1$}
\affiliation{$^1$Physics Department, University of Miami, Coral Gables, Florida FL 33126, U.S.A.\\
$^2$Department of Physics, Chinese University of Hong Kong, Shatin, Hong Kong, China}

\date{\today}

\begin{abstract}
We show that accounting for internal character among interacting, heterogeneous entities generates rich phase transition behavior between isolation and cohesive dynamical grouping. Our analytical and numerical calculations reveal different critical points arising for different character-dependent grouping mechanisms. These critical points move in opposite directions as the population's diversity decreases. 
Our analytical theory helps explain why a particular class of universality is so common in the real world, despite fundamental differences in the underlying entities. Furthermore, it correctly predicts the non-monotonic temporal variation in connectivity observed recently in one such system. 
\end{abstract}


\maketitle

\begin{figure}
\includegraphics[scale=0.40]{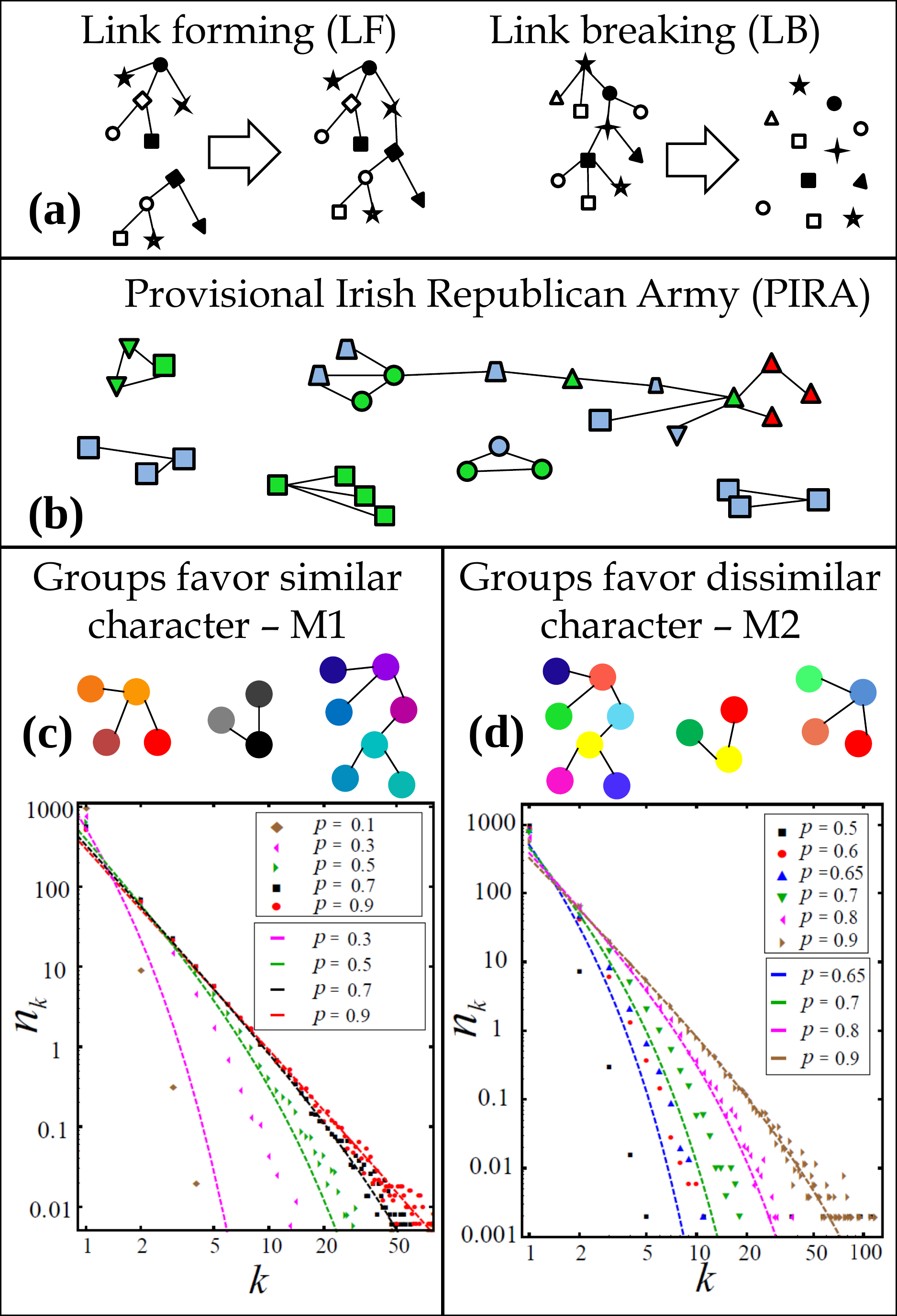}
\caption{\small{(Color online). (a) Our model of interacting characters comprises two types of process: Link formation leading to joining of two groups, and link breaking leading to group fragmentation. (b) Representative portion of PIRA insurgency network in Northern Ireland, adapted from Ref. \cite{B2B}. Different symbols and colors represent different character types (e.g. bomb-maker). It is slightly more connected than our model since all empirical link information is aggregated over a year \cite{B2B}. (c) Groups favoring similar characters (e.g. kin) illustrated by similar colors. Underneath, group size distribution $n_k$ showing simulation (symbols) and analytical (lines) results for different $p$ values. (d) Same as (c) but now for groups favoring diverse characters (e.g. team). }}
\end{figure}

Dynamical grouping underlies myriad collective phenomena across the physical, biological, chemical, economic and social sciences \cite{Korniss,Soulier,Goncalves,palla,redner10,Estrada,JP1,JP2,Song,Caldarelli}. Whether the underlying $N$ objects are particles, people or proteins, the issue of whether they evolve as isolated individuals or aggregates has significant consequences at the macro-level \cite{JP1,JP2,Song,Caldarelli,barabasi,Fortunato,Timme11,dodds,couzin}. Super-radiance is driven by two-level systems coupling coherently via a background boson mode \cite{Acevedo}; many neurodegenerative diseases are driven by aggregation of proteins \cite{zhao09}; large market movements are driven by traders' herding \cite{EZ00,Gabaix,hulst}; insurgencies are driven by informal human groupings \cite{robb,kilcullen,Bohorquez09,SciRep} as are gangs and online guilds \cite{gangs}; brain activity features collective neuronal avalanches \cite{chialvo}; and many-body coherence phenomena are impacted by connectivity within exotic materials \cite{fratini,ourAIP} and networks \cite{Timme11,dodds}.

It is tempting to try applying Physics models of  interacting, identical particles to describe grouping dynamics in living systems. However a serious shortcoming is that the underlying objects (e.g. people, cells, animals etc.) are generally not identical, and it is this heterogeneity that typically dictates their interactions and ultimately their collective behavior. Even simple cells of a given type can have chemical, physical and conformational differences that affect their interactions, while for humans it is usually the characteristics of other group members that dictate whether individuals join or leave a particular group \cite{Uzzi,forsyth}. 
An outstanding question is therefore how this diversity in individual characters affects the dynamics of groups \cite{JP1,JP2,Uzzi,forsyth,Centola,Rodgers}? And how can this individual-level heterogeneity be reconciled with the emergent universality observed across many diverse real-world phenomena? 

This paper attempts to address these questions by adding a simple, continuous `character' variable $x_i$ to each object $i$, and then allowing these characters to influence how objects interact with each other. We assign static $x_{i}$'s randomly from a given distribution $p(x)$ though this will be generalized in future work, e.g. to incorporate experience. 
A single scalar parameter has already been adopted in other contexts within the social science literature \cite{Centola}. Here we restrict our focus to systems where forming links can be costly, e.g. in insurgencies since each link increases the risk of detection \cite{robb,kilcullen,B2B}, in financial trading since new links may leak proprietary information \cite{EZ00,Gabaix}, in the brain since an increase in coordination between neurons will momentarily require additional resources \cite{chialvo}, in coherent many-body states in a fragmented material since each link can increase the chance of a decoherence event \cite{fratini}. Likewise breaking a link in such systems (e.g. through a loss of common purpose, loss of trust, loss of coordination, or loss of coherence respectively) can lead to complete fragmentation of the group (cluster) \cite{caro,EZ00,Gabaix,hulst,robb,kilcullen}. We therefore implement a character-driven fission-fusion mechanism (Fig. 1) that mimics these features, producing sparse networks which are visually similar to those observed empirically (Fig. 1(b)). Previous work \cite{Ruszczycki09,Johnson13,Rodgers} including in the absence of character, suggests that our main conclusions will hold for a variety of model generalizations. 

We define the similarity between objects $i$ and $j$ as $S_{ij}\equiv 1-|x_{i}-x_{j}|$. Though we choose $0\leq \{x_{i}\}\leq 1$, wider ranges do not affect our main conclusions. Objects $i$ and $j$ with similar characters have $S_{ij}$ near unity while those with dissimilar characters have $S_{ij}$ near zero. At each timestep $t$, with probability $p$ an as-yet inexistent link is randomly chosen as a candidate to form. If it forms following the grouping rules based on $S_{ij}$ (see below) it will join together the two groups to which $i$ and $j$ belong (Fig. 1(a)). With probability $1-p$, an existing link is randomly chosen as a candidate to fragment. If it fragments following the grouping rules, the group within which it resides also fragments, mimicking the loss of common purpose, loss of trust, loss of coordination, or loss of coherence mentioned above (Fig. 1(a)).

We first consider each simulation being run using one (and only one) of the following grouping mechanisms. M1: Groups favor similar characters (e.g. kin) as in Fig. 1(c). At a link-forming timestep, the probability that the candidate link actually forms is $S_{ij}$. At a link-breaking timestep, the probability that the candidate link actually breaks is $(1-S_{ij})$. M2: Groups favor diverse characters (e.g. team) as in Fig. 1(d). At a link-forming timestep, the probability that the candidate link actually forms is $(1-S_{ij})$. At a link-breaking timestep, the probability that the candidate link actually breaks is $S_{ij}$. For comparison, we also consider intermediate (M3) and character-free (M4) grouping mechanisms. These are summarized in Table 1.

Figure 2 shows that even for a uniform character distribution $p(x)$, rich behavior emerges. As $p$ increases, the average number of links per object $\left<\lambda\right>$ increases from zero indicating groups spontaneously forming from a population of isolates. Figure 2(b) shows the corresponding rate of change. The position and shape of the onsets depend on the grouping mechanism, with the M2 onset (e.g. team) more abrupt than M1 (e.g. kin) but requiring much higher $p$. This implies that high-diversity groups and teams need to be encouraged by externally establishing a high $p$ ($>p_c$) while kinship groups will naturally arise for almost any $p$. Interestingly, the M1 and M2 onsets are less sharp than the intermediate M3 or character-free M4. This suggests that real-world populations in which character dictates the grouping dynamics, will show far more glassy transitions indicative of frustrated dynamics as compared to the sharp ones in character-free physics models. Results shown are averages over simulations with $N=10^4$ objects, with each simulation comprising $10^5$ timesteps and data collected in the steady state. The SM illustrates results for smaller $N$, and the distribution of similarities $S_{ij}$ for the groups that emerge under mechanisms M1 and M2. 

Our analytical analysis is a mean-field approach, starting with the coupled differential equations for $n_{k}$, the number of groups of size $k$ at timestep $t$ for $k\leq N$: 
\begin{eqnarray}
{\dot n_{k}}&=&-(1-p)Q\frac{(k-1)n_{k}}{\sum_{r=2}^{\infty}(r-1)n_{r}}-2Fp\frac{kn_{k}}{N^{2}}\sum_{r=1}^{\infty}rn_{r}\nonumber\\&+&\frac{Fp}{N^{2}}\sum_{r=1}^{k}rn_{r}(k-r)n_{k-r},\quad k\geq 2\\
{\dot n_{1}}&=&(1-p)Q\frac{\sum_{k=2}^{\infty}k(k-1)n_{k}}{\sum_{r=2}^{\infty}(r-1)n_{r}}-2pF\frac{n_{1}}{N^{2}}\sum_{r=1}^{\infty}rn_{r}
\end{eqnarray}
where $F$ is a mean-field probability of a link being formed between two randomly chosen objects, while $Q$ is a mean-field probability that an arbitrarily chosen link will break and hence that group will fragment. Since our focus is on networks that are naturally sparse \cite{robb,kilcullen}, we take a group of size $k$ as having $(k-1)$ essential links \cite{robb,kilcullen} in Eq. (1), though any number $O(k)$ would generate similar conclusions. In the steady-state, these equations yield two possible solutions for the number of isolated individuals (see SM): $n_1 = N$ or $n_1 = \frac{pF +
 (1-p)Q}{2pF}N$.
Since $n_1\leq N$, a transition will arise when $[pF + (1-p)Q] = 2pF$ from a population comprising $100\%$ isolates to one with cohesive groups, i.e. at the critical point 
\begin{equation}
p_c = Q(F+Q)^{-1} \ .
\end{equation}
For $p > p_c$, each $n_{k}$ for $k\geq 2$ changes from zero to 
the exact expression 
\begin{equation} \label{eq-n_k}
 n_k = |  \frac{1}{2}!\  [2k\gamma (k!) (\frac{1}{2} - k)!]^{-1}\ 
            (4 \gamma n_1)^k |
\end{equation}
where
\begin{equation}
\gamma = pF (N-n_1)(N [Q(1-p)N+2pF(N-n_1)])^{-1}\ .
\end{equation}
We can evaluate $F$ and $Q$ analytically to obtain $p_{c}$ for grouping mechanisms M1-M4: For a uniform character distribution $p(x)$, the probability density function (PDF) $f(y)$ of the similarity $y=S_{ij}$ is given by $f(y)=2y$, with $y\in [0,1]$. For mechanism M1, the probability $F$ that two objects will be linked is $\int_{0}^{1}f(y)ydy=2/3$. Similarly, the PDF of $y$ associated with links is $g(y)=3y^2$, hence the probability $Q$ that a randomly selected link breaks is $\int_{0}^{1}g(y)(1-y)dy = 1/4$. 
Following this procedure for M1-M4 yields the theoretical values shown in Table 1. 

\begin{table*}
\centering
\begin{tabular}{ p{3.6cm} p{2.5cm} p{2.5cm} p{4.0cm} p{2.0cm} p{1.5cm} p{1.5cm} }
\hline
 \centering  & \centering Link forming probability  & \centering Link breaking probability &  \centering$p_{c}$ (mean field theory)& \centering$p_{c}$(numerical)& \centering$F$ &$Q$ \\ [0.5ex] 
\hline \hline
\centering M1 (e.g. kinship) &   \centering$S_{ij}$ & \centering$1-S_{ij}$ &  \centering$3/11$ &  \centering$0.10$ &  \centering$2/3$ &  $1/4$ \\ [0.8ex]
\centering M2 (e.g. team) &  \centering$1-S_{ij}$ & \centering$S_{ij}$ &  \centering$3/5$ &  \centering$0.51$ & \centering$1/3$ & $1/2$ \\[0.8ex]
\centering M3 intermediate &   \centering$S_{ij}$ & \centering$S_{ij}$ & \centering$9/17$ & \centering$0.49$ &  \centering$2/3$ & $3/4$ \\ [0.5ex] 
\centering M4 character-free &  \centering$1$ & \centering$1$ & \centering$1/2$ & \centering$0.50$ &  \centering$1$ & $1$ \\[0.8ex]
\hline
\end{tabular}
\caption{Different grouping mechanisms M1-M4. $F$ and $Q$ calculated analytically for uniform character distribution $p(x)$. Mean-field result $p_c=Q(F+Q)^{-1}$.}
\label{table:1}
\end{table*}%


\begin{figure}
\centering
\includegraphics[scale=0.55]{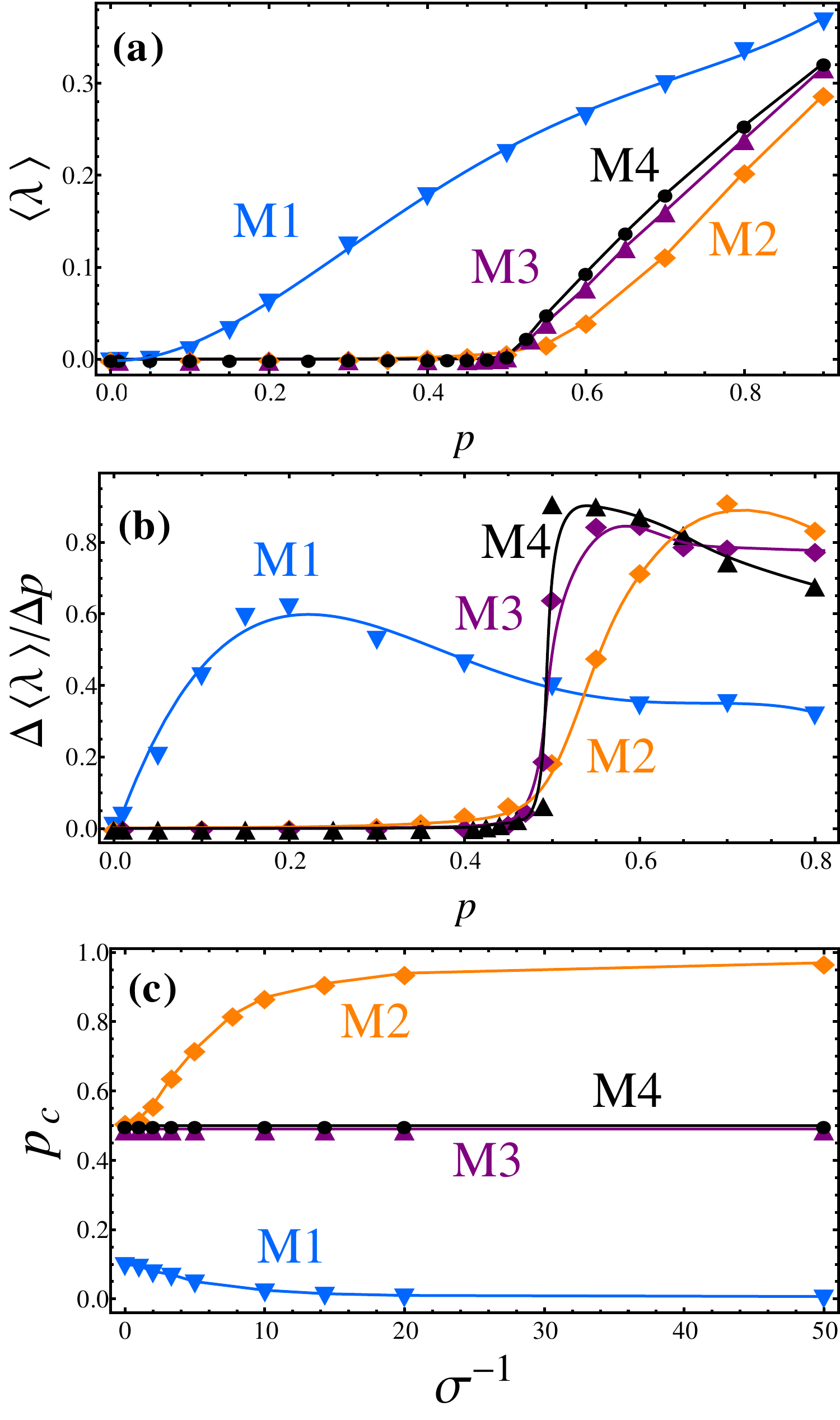}
\caption{\small{(Color online) (a) $\left<\lambda\right>$ versus $p$ for $N=10^4$ objects. Blue points: grouping mechanism M1 favoring similar characters (e.g. kinship). Orange points: grouping mechanism M2 favoring diverse characters (e.g. team). Purple points: M3 intermediate between M1 and M2. Black diamonds: M4 character-free. Table 1 shows comparison to theoretical $p_c$. (b) Rate of change. (c) $p_c$ for M1 (bottom, blue), M2 (top, orange), M3 and M4 (horizontal) versus inverse standard deviation |($\sigma^{-1}$) of population's character distribution $p(x)$.}}
\end{figure}

\begin{figure}
\centering
\includegraphics[scale=0.45]{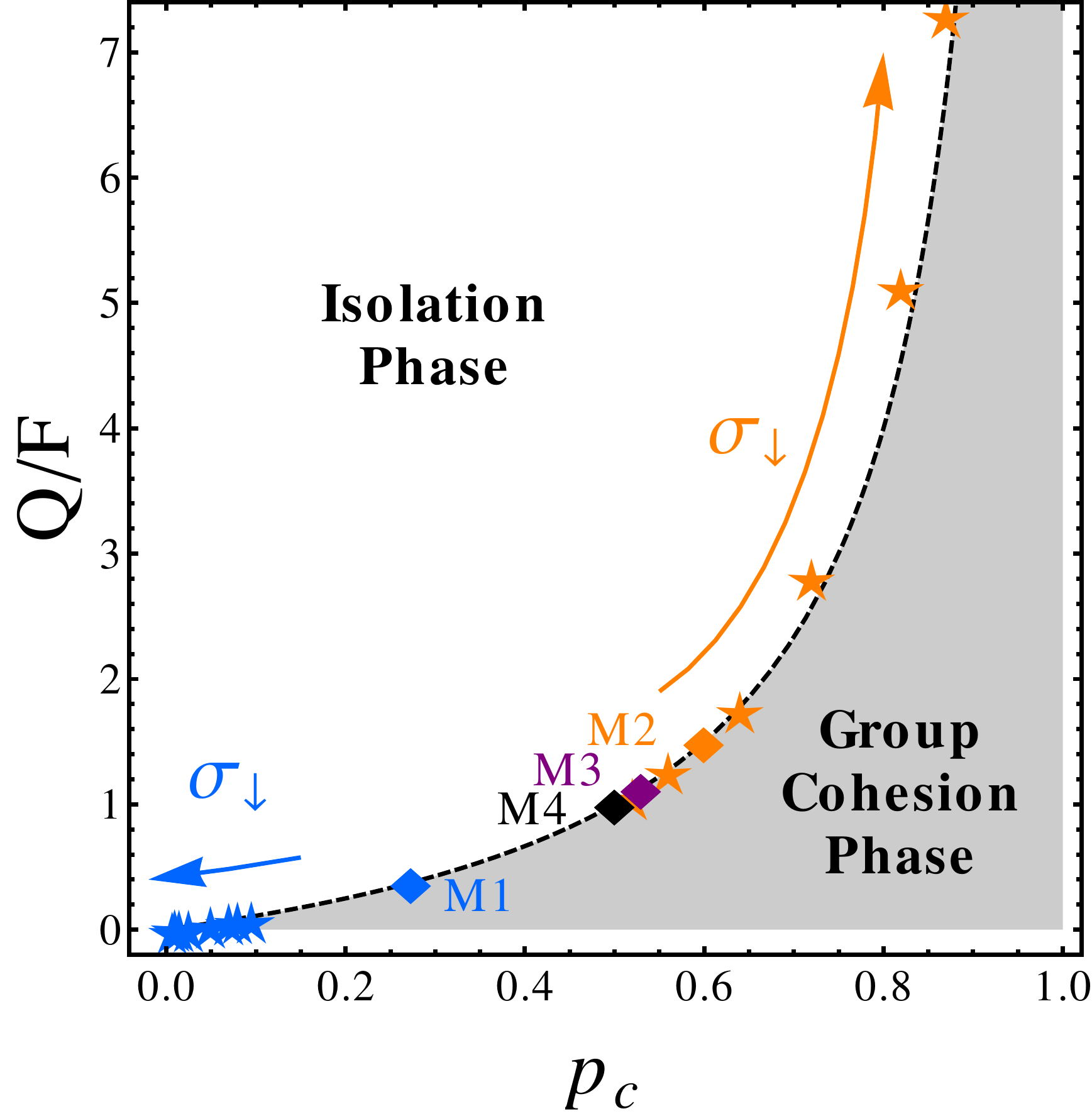}
\caption{\small{(Color online) Phase diagram. Curved phase boundary is our mean-field analytical result $p\equiv p_c=Q(F+Q)^{-1}$, i.e. $Q/F=p_c(1-p_c)^{-1}$. Diamonds show $p_c$ for uniform character distribution $p(x)$. Stars show numerical results for gaussian $p(x)$ from Fig. 2(e). M1 blue, M2 orange.}}
\end{figure}

Figures 1(c)-(d) and Table 1 show good agreement between numerical simulation and our mean-field theory for $\{n_k\}$ and $p_c$. Differences are due to neglect of higher-order correlations. Equation (4) further reduces (see SM) to the approximate form for $p>p_c$:
\begin{equation}
n_k = \frac{N}{2\sqrt\pi}\frac{p(1-p_c)}{p-p_c} \left[1-\left(\frac{p_c(1-p)}{p(1-p_c)}\right)^2\right]^k\ k^{-5/2} \ .
\end{equation}
Equation (6) predicts an approximately exponential cutoff at high $k$ that depends on the grouping mechanism through $p_c=Q(F+Q)^{-1}$, together with a $5/2$ power law exponent that does not. As data from real-world systems improves, it should be possible to estimate $p_c$ and $p$, and hence $Q/F$, to infer likely character-driven grouping mechanisms in a given system. 
The $5/2$ exponent is exactly that observed empirically for (i) the severity of attacks inflicted by insurgent groups on a civilian population, indicating the size distribution of the insurgent groups \cite{Bohorquez09,SciRep}; (ii) the distribution of stock transaction sizes, indicating the herd sizes of similar-minded traders \cite{Gabaix}; (iii) the size distribution of neuronal avalanches, given avalanche initiation by a randomly chosen neuron (i.e. $k.k^{-5/2}\equiv k^{-3/2}$ \cite{chialvo}); (iv) the size distribution of pockets of superconducting coherence in fragmented materials \cite{fratini}. It is also close to the two values of 2.3 obtained from a size study of 100 gangs in Chicago, and separately in Manchoukuo in 1935 \cite{richardson}.

Figure 2(c) shows that $p_{c}$ shifts in opposite directions for M1 and M2 as the heterogeneity  of the underlying population is reduced, using a gaussian character distribution $p(x)$ with mean $\mu=0.5$ and standard deviation $\sigma$. This implies that teams require an ever higher $p$ to form as a population becomes more homogeneous, with the population eventually comprising completely isolated individuals for all $p$. By contrast, kin groups require an ever lower $p$ to form. Figure 3 shows the phase diagram. The numerical simulation results lie remarkably close to the analytic curve $Q/F=p_c(1-p_c)^{-1}$, providing further support for our mean-field analysis. 

Though PIRA (Fig. 1(b)) is the best-known insurgency network to date \cite{B2B}, the data is still unfortunately insufficient to infer the actual grouping mechanism since links are aggregated over years, which is why it appears more dense than snapshots of our model. However we can test our character model against the recent state-of-the-art case study \cite{B2B} that suggests that PIRA underwent a bottom-up transition over time from a rather homogeneous organization toward team-like structures facilitated through a process of individual contact. We start our model PIRA population with an M1 grouping mechanism favoring similar character links. An individual is introduced who uses an M2 grouping mechanism favoring diverse character links and hence favoring team formation, and who is able to spread its use to anyone with whom he/she instantaneously shares a cluster. They then become spreaders (susceptible$\rightarrow$infected) reflecting the fact that team-like structure became recognized as improving PIRA's operational efficacy and hence got reinforced over time by contacts at grass-roots level. Figure 4 shows the resulting prediction concerning connectivity from our model and the actual PIRA data. The same non-monotonic dynamics arise in both.

\begin{figure}
\centering
\includegraphics[scale=0.35]{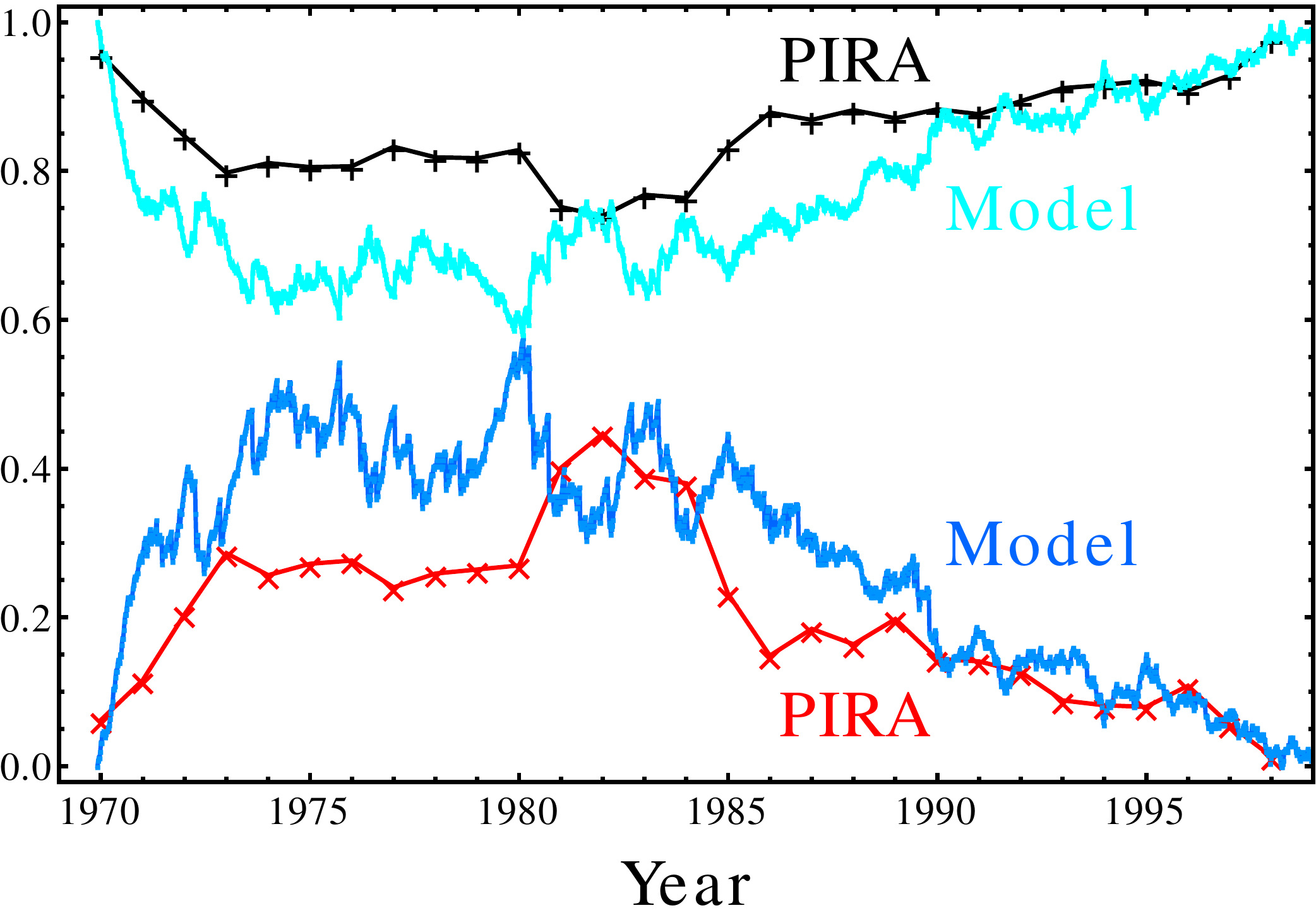}
\caption{\small{(Color online) Our model's prediction vs. actual PIRA temporal variation for (top two curves) the fraction of isolated individuals, and (bottom two curves)  
the ratio between the total number of network links and the total number of individuals. PIRA data adapted from Ref. \cite{B2B}.}}
\end{figure}

In summary, we have shown that rich phase transition dynamics emerge when the objects in a population possess an internal character variable. Our analytical theory explained why a particular statistical universality is so ubiquitous in real-world systems, despite fundamental differences in the composite objects and their interactions. Our findings open a path toward understanding how different grouping mechanisms (e.g. M1 vs. M2) affect diseases or memes spreading in realistic (i.e. heterogeneous) populations.  In physical systems, the different grouping mechanisms (e.g. M1 vs. M2) can be used to mimic the tuning of particle-particle interactions in an exotic material, with $p$ acting like an inverse temperature. 

\begin{acknowledgments}
We are grateful to Chaoming Song and Stefan Wuchty for discussions and to John Horgan and Paul Gill for sharing their PIRA data and specialist knowledge. 
\end{acknowledgments}

\end{document}